There are three reasons that have inspired me to translate Frenkel's article [1]. First, although the article has been published decades ago (1948), there is still no English translation. Second, while I was attending the 57th APS DFD conference, I have realized that Frenkel's results have attracted great interest by the attendees, however the general community is widely unaware of his elegant findings. And third, when I have read papers that cite Frenkel's publication (I have only found four [2-5]), I have realized that many people misunderstand his result. For example, J.J. Bikerman 1950 [2], like others after him, has only understood Frenkel's conclusion superficially, considering only the *scenario* of the drop motion, instead of the entire mathematical subtlety that the problem involves. In my opinion, the beauty of the Frenkel's article is in the details of his statements.

Y.I. Frenkel was probably the first person to state and solve both problems of capillarity: the problem concerning the droplet shape on the horizontal *and* the inclined surface. He pointed out that the Young-Neumann formula plays the role of a boundary condition, and it needs to be generalized when the droplet surface exhibits a contact angle hysteresis. He has shown that the formula $l\sigma(\cos\theta_1 - \cos\theta_2) = gm\sin\alpha$ (that people studying the motion of the droplet on the inclined surface due to experiment find so lovely) is just a boundary condition; it is not universal, and it depends on the statement of the problem.

I am indebted to Vera Sazonova and Patrick Weidman for their great help in preparation of this translation.

1. Y.I. Frenkel, *J. Exptl. Theoret. Phys.* (USSR), **18**, 659, 1948
2. J.J. Bikerman, *J. Colloid Science*, **5**, 349, 1950.
3. C.G.L. Furmidge, *J. Colloid Science*, **17**, 309, 1962.
4. C.W. Extrand, A.N. Gent, *J. Colloid and Interface Science*, **138**, 431, 1990.
5. P. Roura, J. Fort, *Phys.Rev.E*, **64**, 011602, 2001.

Dr. Viatcheslav Berejnov
Cornell University
Physics Department, Clark Hall, B5
Ithaca, NY, 14853-2501
berejnov@cornell.edu

*J. Exptl. Theoret. Phys.* (USSR), **18**, 659, 1948 – this is ref. of the original Russian variant# On the behavior of liquid drops on a solid surface
# 1. The sliding of drops on an inclined surface
*Y.I. Frenkel*

translated by Berejnov Viatcheslav, *berejnov@cornell.edu*, 2005

**Abstract**

The problem of contact angle variations of a droplet that is at rest on an inclined solid surface is considered. The limit inclination angle at which the droplet begins to move is determined. It is found that the motion of the drop takes place due to the transport of its own liquid forward with a simultaneous rear edge detachment from the solid surface. In the case of a good wetting, the drop leaves an unstable liquid trace behind, which in time ruptures into a number of small droplets.**1. Contact angles.**

For mathematical simplification of some capillarity problems, and particularly for the contact angle problem of a drop on a solid surface, it is useful to approximate the shape of the drop with an infinite cylinder. This simplification allows us to reduce the problem from three to two dimensions. In this case, the "drop" should be considered as a two-dimensional figure with contour $s$ (partly constituted by the surface of the solid) and characterized by the conservation of its area $S$. In addition, when the surface is horizontal the drop has an equilibrium contact angle $\theta_0$, which according to Neumann's law is given by:

$$(1) \quad \cos\theta_0 = (\sigma_{10} - \sigma_{21})/\sigma_{20}$$

where $\sigma_{20}$, $\sigma_{21}$, $\sigma_{10}$ are surface tension coefficients on the liquid/air boundary, on the liquid/solid boundary, and on the solid/air boundary, respectively (Fig. 1).

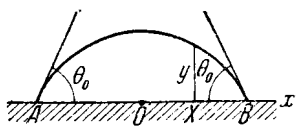

Рис. 1

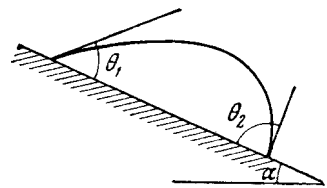

Рис. 2

On the inclined solid surface the contact angle varies and reaches a value $\theta_2 > \theta_0$ for the front angle and $\theta_1 < \theta_0$ for the rear angle, respectively (Fig.2[1]). As the inclination angle increases the angle $\theta_2$ increases and the angle $\theta_1$ decreases. For some critical angle $\alpha = \alpha^*$ which corresponds to values of $\theta_1 = \theta_1^*$ and $\theta_2 = \theta_2^*$, the drop begins running down the surface (for sufficiently large drops). In the case of poor wetting, the drop does not leave any liquid trace behind as it does in the case of good wetting. This liquid trace can in time

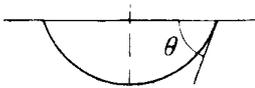

Рис. 3

separate into the small droplets due to its instability (similarly to the instability of a three-dimensional liquid jet).

The small droplets are pinned by the solid surface even for $\alpha = 180^O$ which corresponds to the hanging drop case (Fig. 3). This last example illustrates that the Neumann law is not general and the contact angle can vary due to external forces that are acting on the drop. Another evidence of Neumann law limitations is the phenomena of rupturing of the liquid trace behind the drop into separate droplets and a number of effects known as "contact angle hysteresis" which even today are still puzzling from the theoretical point of view (*written in 1948 – B.V. note*).

---

[1] Figure 2 is modified by Photoshop to eliminate the typographical errors presented on the original. (*B.V.*)

## 2. The shape of a droplet lying on a horizontal surface.

Before considering the phenomena mentioned above, we will give a more general definition of "contact angle" proceeding from the principle of minimum total energy for the drop on a flat horizontal surface.

This energy consists of the following parts:

1. The surface energies $U_1 = (\sigma_{12} - \sigma_{10})s_1$ corresponding to the fluid/solid interface and $U_2 = \sigma_{20}s_2$ corresponding to the free part of the droplet's surface. Let $s_1$ be the length of the AB interval (Fig. 1) and $s_2$ be the length of the curve connecting A and B. We calculate the energy of the drop per unit length of the cylinder whose normal section is the two-dimensional drop we are considering. The fact that $U_1$ is proportional to the difference between $\sigma_{12}$ and $\sigma_{10}$ corresponds a reduction of the free *area*[2] of the solid as $s_1$ increases.

2. The gravitational energy (or an energy of other external forces that are acting on the drop; the inertia force, wind pressure, etc.) is $U_3 = \frac{1}{2}\rho g \int y^2 dx$. This energy again is given per unit length of the cylinder that is used instead the real drop. $X$ is the abscissa of a certain point, $y$ is the height of the drop contour under this point, $\rho$ is the fluid density, and $g$ is the gravitational acceleration constant. Therefore, the total energy of the drop is given by the formula:

$$(2) \quad U = \int \left[ (\sigma_{12} - \sigma_{10}) + \sigma_{20}\left(\sqrt{1 + y'^2} + \frac{y^2}{2a^2}\right) \right] dx$$

where $y' = \frac{dy}{dx}$, $\sqrt{1 + y'^2}\,dx = ds_2$ is the differential element of the drop contour and $a^2 = \frac{\sigma_{20}}{\rho g}$ is a parameter with dimensions length squared (in case of water a ≈ 0.3 cm).

To find the minimum of $U$ we should vary both the function $y(x)$ and the limits of variation of the independent variable $x_{\max} = X$ and $x_{\min} = -X$. Setting the variation of $U$ to zero leads to the vanishing of the difference between the integrand $L$ and the function $y'\partial L/\partial y'$ on the both boundaries, respectively, i.e. it leads to the equality[3]

---

[2] I believe that instead of the *area* here should be the *energy*. The size of the area has nothing to do with the $\sigma_{12}$ and $\sigma_{10}$ difference, it just proportional to *s*, and it makes the sentence trivial and completely independent of $\sigma_{12}$ -$\sigma_{10}$. However, the free energy depends on such difference in a way as described in the sentence because of the negative sign ($\sigma_{12}$<$\sigma_{10}$). (*B.V.*)

[3] Here and later in the article $\sigma_{20} = \sigma$. (*B.V.*)

$$(\sigma_{12} - \sigma_{10}) + \sigma\left(\sqrt{1+y'^2} - \frac{y'^2}{\sqrt{1+y'^2}}\right) = 0$$

or

$$(3) \quad (\sigma_{12} - \sigma_{10}) + \frac{\sigma}{\sqrt{1+y'^2}} = 0, \quad (x = \pm X)$$

since at $x = \pm X$, $y=0$. Furthermore, the integral $\int_{-X}^{+X} \delta\left(\sqrt{1+y'^2} + \frac{y^2}{2a^2}\right)dx$ does not appear because of the invariance of the drop's "surface"[4] $S = \int y dx$. This invariance condition can be taken into account by adding the integral $\lambda \delta S = \lambda \int \delta y dx$ to the previous integral, where λ is a Lagrange multiplier. Therefore, the extremal drop profile is defined by condition (3), which agrees with the Neumann law (1) for the contact angle for $x = \pm X (\sqrt{1+y'^2} = \frac{1}{\cos\theta_0})$, and with the differential Euler equation:

$$(d/dx)\partial L/\partial y' = \partial L/\partial y$$

where

$$L = \sqrt{1+y'^2} + \frac{y^2}{2a^2} - \lambda y$$

i.e. $\quad (4) \quad \dfrac{d}{dx}\dfrac{y'}{\sqrt{1+y'^2}} = \dfrac{y}{a^2} - \lambda.$

The expression on the left hand side is the curvature of the drop contour at the point, or the inverse value of the radius of the curvature R.

### 3. An approximate solution for perfect wetting.

The integration of the equation (4) is not difficult since it does not contain the independent variable of x. However, for simplification, and to clarify the physical meaning of the constant λ, we first consider the linearized version of (4):

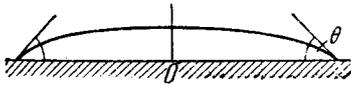

Рис. 4

$$(5) \quad \frac{d^2 y}{dx^2} = \left(\frac{y}{a^2}\right) - \lambda,$$

The general solution of this equation is given by:
$$y = A_1 e^{x/a} + A_2 e^{-x/a} + \lambda a^2.$$

Because of the symmetry of the drop's profile $y(x)$ with respect to the center ($x=0$) it is possible to assume that $A_1 = A_2 = -A$, which leads to:

$$(6) \quad y = \lambda a^2 - 2A\mathrm{ch}\left(\frac{x}{a}\right).$$

---

[4] Instead of "surface" I guess the more appropriate term here is 2-d volume. Because by the sense this integral gives an area of the 2-d drop that remains constant for any deformation of the drop shape. (*B.V.*)

Linearization of equation (4), or its replacement by equation (5), is possible only in the case of *perfect wetting* of the solid by the liquid. The drop profile is then an inverted catenary (see Fig. 4).

The area $S = \int y dx$, using (6), is expressed by the formula $S = 2\lambda a^2 X - 4A\text{sh}(X/a)$. Hence, the expression for the constant $\lambda$ is:

$$(6a) \quad \lambda = \frac{S}{2a^2 X} + \frac{2A}{a^2}\frac{\text{sh}(X/a)}{X/a}$$

The coefficient A can be determined from the condition $y=0$ for $x=\pm X$, i.e. it is given by the formula:

$$\lambda a^2 = 2A/\text{ch}(X/a)$$

or

$$(6b) \quad A[\text{ch}(X/a) - (a/X)\text{sh}(X/a)] = \frac{S}{4X}.$$

For $y=0$ we have

$$(y')_{x=+X} = 2(A/a)\text{sh}(X/a) = \text{tg}\,\theta_0 \approx \theta_0.$$

Hence, for given values of $\theta_0$ and $S$, $X$, which is the length ("area") of the solid/liquid interface, is defined uniquely. Notice, that this length is constrained by the maximum height of the drop $h = y_{\max} = \lambda a^2 - 2A$. These results hold in the general case of a flat droplet with arbitrary curvature.

Rewriting the expression $(d/dx)(y'/\sqrt{1+y'^2})$ in equation (4) in the form

$$\frac{y''}{(1+y'^2)^{3/2}} = \frac{y'dy'/dy}{(1+y'^2)^{3/2}} = -\frac{d}{dy}\frac{1}{\sqrt{1+y'^2}},$$

we arrive at

$$(1+y'^2)^{-1/2} + \left(\frac{y^2}{2a^2}\right) - \lambda y = C,$$

where $C$ is a constant. It is possible to define $C$ from the condition $y'=0$ in the center of the drop at $y=y_{\max}=h$. This yields:

$$C = \left(\frac{h^2}{2a^2}\right) - \lambda h + 1$$

and consequently:

$$(7) \quad (1+y'^2)^{-1/2} = \left(\frac{1}{2a^2}\right)(h^2 - y^2) - \lambda(h - y) + 1.$$

Since $(1+y'^2)^{-1/2} = \cos\theta$, where $\theta$ is an angle between the tangent of the drop contour and $x$-axis, it is possible to rewrite this formula in the form

$$(7a) \quad \sin^2(\theta/2) = \frac{1}{2}\lambda(h-y) - \frac{1}{4a^2}(h^2 - y^2).$$

Let us define for brevity the right part of the expression (7) as $f(\lambda,y)$. Therefore, we have:

$$S = \int y dx = \int y(dx/dy) dy = \int (y/y') dy$$

or

(8) $$S = \int \frac{yf(y,\lambda)}{\sqrt{1-f^2(y,\lambda)}}dy.$$

In general it is possible to find the parameter $\lambda$ from the last equation. The equation for the drop contour can be derived from the equation (3) by solving for $y'$. Therefore, we find:

(8a) $$x = \int (1-f^2)^{-1/2} f dy.$$

## 4. Variation of drop contact angles on a tilted surface at equilibrium.

Let us now consider the drop behavior on the inclined surface. We should only change the expression for the potential energy $U_3$ while keeping the same coordinate system ($x$ is along the surface, $y$ is perpendicular to it). Let us introduce an additional coordinate system, which consists of the horizontal axis $\xi$ and vertical axis $\eta$. We denote by $\alpha$ the angle between coordinate systems $(x, y)$ and $(\xi, \eta)$. We calculate the energy $U_3$ relative to any fixed horizontal plane, for example one containing the axis $\xi$. We assume both coordinate frames $(x, y)$ and $(\xi, \eta)$ have the same origin $x=0$, $y=0$ which coincides with the drop center at $\alpha=0$. The height of any point on the drop's profile with respect to the $\xi$ axis is $\eta = y\cos(\alpha) - x\sin(\alpha)$ (this value can be both positive or negative). Therefore,

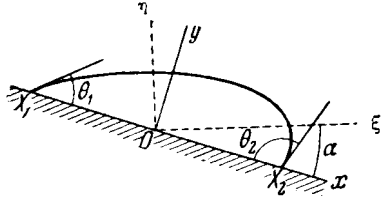

Рис. 5

$$U_3 = \rho g \iint [y\cos(\alpha) - x\sin(\alpha)] dx dy$$

or

$$U_3 = \frac{1}{2}\rho g \int [y\cos(\alpha) - x\sin(\alpha)] y dx.$$

In the last expression, the integration is performed over the free part of the drop's surface. Thus, for the total energy we have:

(9) $$U = \int_{X_1}^{X_2} \left\{ (\sigma_{12} - \sigma_{10}) + \sigma_{20}\left[\sqrt{1+y'^2} + \frac{1}{a^2}\left(\frac{y^2}{2}\cos(\alpha) - xy\sin(\alpha)\right)\right]\right\} dx.$$

Using the fact that $\int y dx = S = const$ this expression corresponds to the following Euler equation

(10) $$a^2 \frac{d}{dx}\frac{y'}{\sqrt{1+y'^2}} = y\cos(\alpha) - x\sin(\alpha) - \lambda a^2.$$

The boundary conditions are different for the top ($X_1$) and bottom ($X_2$) drop edges. If we assume $\delta X_1 = \delta X_2 = \delta X$ which corresponds to a virtual downward shift of the drop as a whole, then the energy variation $(\sigma_{12} - \sigma_{10})(x_1 - x_2)$ is equal to zero and the boundary condition is

(11) $$\sigma(\cos(\theta_1) - \cos(\theta_2)) = gm\sin(\alpha).$$

This condition describes the equilibrium between the lengthwise force components of surface tension and gravity. In this case, the front angle $\theta_2$ grows and rear angle $\theta_1$ is decreases relative to the angle $\theta_0$, which corresponds to a horizontal surface (in the case $\theta_2 > \pi/2$, $-\cos(\theta_2) = |\cos(\theta_2)|$. The parameter $m = \rho S$ denotes the mass of the cylindrical drop per unit length.

To determine the angles $\theta_1$ and $\theta_2$ separately, it is necessary to add one more condition to equation (11). This new condition should correspond to such variation of $X_1$ and $X_2$ for which the integral $\iint (y\cos(\alpha) - x\sin(\alpha))dxdy$ is constant. This problem is reduced to a generalization of the Neumann formula (1) to the case of an inclined plane. Unfortunately, we have not yet solved it theoretically. However, our experiments show with considerable accuracy that it is possible to assume:

$$(12) \qquad \theta_1 = \theta_0 - \Delta\theta, \quad \theta_2 = \theta_0 + \Delta\theta.$$

Taking into account (11) we have:

$$(12a) \qquad 2\sigma \sin(\theta_0)\sin(\Delta\theta) = gm\sin(\alpha).$$

To illustrate the formulas above, we consider the linearized equation (10), i.e.

$$(13) \qquad a^2 y'' = y\cos(\alpha) - x\sin(\alpha) - \lambda a^2.$$

The solution of this equation is:

$$y = A\exp\left(\frac{x}{a}\sqrt{\cos\alpha}\right) + B\exp\left(\frac{x}{a}\sqrt{\cos\alpha}\right) + x\operatorname{tg}\alpha + \frac{\lambda a^2}{\cos\alpha}$$

or

$$(14) \qquad y = \frac{\lambda a^2}{\cos\alpha} - A\operatorname{ch}\left(\frac{x}{a}\sqrt{\cos\alpha}\right) + x\operatorname{tg}\alpha.$$

Letting the roots of the right hand side be $X_1$ (<0) and $X_2$ (>0) and the corresponding derivatives be $y'_1$ and $y'_2$, respectively, we obtain:

$$y'_1 = -\frac{A}{a}\sqrt{\cos\alpha}\ \operatorname{sh}\left(\frac{X_1}{a}\sqrt{\cos\alpha}\right) + \operatorname{tg}\alpha = \operatorname{tg}\theta_1,$$

$$y'_2 = -\frac{A}{a}\sqrt{\cos\alpha}\ \operatorname{sh}\left(\frac{X_2}{a}\sqrt{\cos\alpha}\right) + \operatorname{tg}\alpha = \operatorname{tg}\theta_2.$$

The maximum height of the droplet, $y_{\max} = h$, is defined by the expression:

$$-\frac{A\sqrt{\cos\alpha}}{a}\operatorname{sh}\left(\frac{X}{a}\sqrt{\cos\alpha}\right) + \operatorname{tg}\alpha = 0$$

and it is shifted to the positive direction of the $x$ axis.

## 5. The limit of the inclination angle and the mechanism of drop movement.

If a drop is large enough, then at a certain inclination angle $\alpha=\alpha^*$ it begins to run down the surface, not leaving any visible trace behind in the case of poor wetting, or leaving a liquid trace in case of perfect wetting.

Before discussing this effect, we should note the known fact from hydrodynamics that a fluid flowing over a solid sticks to its surface, i.e. the fluid cannot slip with respect to the surface. Therefore, the only mechanism for the drop to run down is a "pouring" of the fluid from the rear edge of the drop to the front edge. This mechanism is similar to the motion of the caterpillar tractor (Fig. 6).

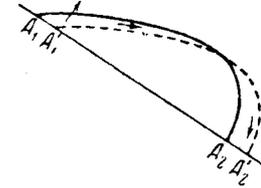

Рис. 6

Such "pouring" begins only when the work of the gravitational force is greater than or equal to the work of adhering forces of the drop's rear edge.

To tear off the rear edge of the (two-dimensional) drop from the surface on the segment of length $A_1A_1'=dx$, it is necessary to apply work in the amount $\Delta\sigma \cdot dx_1$, where $\Delta\sigma = \sigma_{20} + \sigma_{10} - \sigma_{12}$ is the energy of adhesion of the fluid to the solid per unit area. Due to the process of pouring of the detached layer of the fluid from the rear edge to the front edge, the drop is shifted downwards with the same profile by a segment $dx_2=dx_1$. The gravity force does the work:

$$gm \sin\alpha \, dx_1 \,^5.$$

Therefore, the limit of the inclination angle $\alpha=\alpha^*$ for which the drop begins to run down the plane is determined by the expression:

(15) $\qquad \Delta\sigma = gm \sin\alpha^*$.

Substituting for $gm\sin\alpha^*$ the expression $\sigma(\cos\theta_1^* - \cos\theta_2^*)$, where $\theta_1^*$ and $\theta_2^*$ are the limiting values of the contact angles, we have

(15a) $\qquad \Delta\sigma = \sigma(\cos\theta_1^* - \cos\theta_2^*)$.

Let us note that the maximum value of the right hand side is equal to $2\sigma$. If the adhesion energy $\Delta\sigma$ is greater than the self-adhesion energy of the fluid, then the drop does not detach from the solid interface. Instead, the drop leaves behind a liquid trace approximately $10^{-5}$ cm thick (the order of the surface force length scale), as if making a pathway for itself. In that case, expression (15) should be written:

(15b) $\qquad 2\sigma = gm \sin\alpha^*$.

It is clear from the last expression that it is not possible for the drop to roll down when $m < 2\sigma/g$ (i.e. $\sin\alpha^* \leq 1$). In that case, the droplet stays in place even if the surface is vertical or even more, if the surface is overturned (see Fig. 3 above).

Substituting $m$ by $\pi R^2 \rho$, where $R$ is the drop's radius (or rather the radius of the cylinder) in the non-deformed state, we arrive at the minimal value of $R$ for which the drop begins to move (at $\alpha^* = \pi/2$):

$$R = a\sqrt{2/\pi}.$$

---

[5] The fact that instead of the area uncovered near the drop's rear edge, the drop covers the same area on the front edge (*Fig. 6 B.V.*) has no special importance for determination of the drop motion conditions. The gravity forces that lead to tearing of the liquid from the drop's rear edge are weaker than the adhesion forces. However, the primary importance lies in the product of the force by the arm, as it is in case of the solid arm.

Because of the practical importance of the above relations, we will reconsider them for the case of a three-dimensional drop. Suppose the drop contact with the horizontal solid forms a circle of radius *r*. When the drop moves a distance $dX_1$, it reveals at the rear edge of the drop the surface of a crescent enclosed by the initial and final position of the circle (Fig. 7). The area of this crescent is equal to

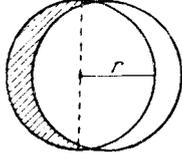

Рис. 7

$$dX_1 \int r \sin\phi \, d\phi = 2r \, dX_1 .$$

Equating the product of the last expression and $\Delta\sigma$ to the gravity force work on the distance $dX_1$ gives:

(16) $\quad 2r\Delta\sigma = mg \sin\alpha^*.$

Of course this formula is true in case of poor wetting only (i.e. in the case $\Delta\sigma < 2\sigma$, for example water on paraffinized glass). In the case of good wetting, $\Delta\sigma$ in the last expression should be replaced by $2\sigma$. Rewriting *m* as $(4\pi/3)\rho R^3$, where *R* is a radius of the undeformed drop, gives:

(16a) $\quad \sin\alpha^* = 3a^2 r / \pi R^3,$

where $a^2 = \Delta\sigma/2\rho g$ (or $\sigma/2\rho g$ for $\Delta\sigma > 2\sigma$). In case of water $(\sigma/2\rho g)^{1/2} = 0.3$ cm, so if $r \sim R = 0.3$ cm then $\sin\alpha^* = 1$, and consequently the drop is not able to move.

The above discussion on the drop's motion over the solid can be applied with minor variations to motion of gas bubbles on a solid-liquid interface (for example on the surface of a carbon electrode during the electrolysis of aluminum from $AlF_3$). The only differences are that in the case of poor wetting ($\theta < \pi/2$) the bubble has the same shape as the drop would in the case of good wetting and vice versa, and that the bubble moves upwards along the solid whereas the droplet moves downwards. The shape of the bubble changes only slightly under hydrostatic pressure compared to the drop's shape, so that the bubble on top of a horizontal solid surface has the same shape as a drop hanging from an overturned solid surface.

**6. Contact angle hysteresis and similar phenomena.**

The liquid trace left behind the running drop, in the case of perfect wetting, should have a flat jet shape i.e. the shape of a film with the slowly decreasing width (as the drop size decreases slowly during motion). This film is presumably unstable in the same way as the usual three-dimensional jet which, according to Rayleigh, should separate into distinct droplets. If the liquid wets the solid only partially (for example, a drop on a slightly dusted glass surface), then in time the liquid trace of the running down drop separates into small droplets. These droplets are elongated along the general direction of drop's motion and form a dash-like trace on the surface. These droplets are very well known and can be observed on the window of a fast moving train or car during rain. The worse the wetting, the less time it takes for separation into droplets.

To determine the "liquid trace" thickness it is necessary to take into account the interaction between the liquid and the solid in a less schematic fashion than above. For that we should not consider the capillary forces as being constrained to the interface, but as creating some surface field of interaction near the solid. The range of that field is on the order of $10^{-5}$ cm and its strength is inversely proportional to the cube of the distance.

The average thickness of the liquid film left on the solid surface should reach the same order of magnitude estimation.

The rupture of the liquid film into separate droplets is similar to the rolling up process of the previously over-spread drop, or to the inverse process of the spreading of a drop that was previously over-rolled. In the last two cases (horizontal surface), the deviation of the contact angle $\theta$ from its common value $\theta_0$ that corresponds to the minimum energy governs the processes. The contact angle $\theta$ approaches $\theta_0$ due to the change in the area of the droplet-solid interface. Moreover, similarly to the case of a downward running drop, the drop is not really sliding on the surface. The mechanism of the drop's spreading and rolling up is comprised of the motion of the fluid constituting the drop from the bulk (center and top of the drop) to the drop's edge or, inversely, from the drop's edge (bottom) to the bulk (drop center).

For example, if at the beginning the drop has an over-spread shape (solid line in Fig. 8) and tries to retract to a less spread shape (dash line in Fig. 8), then the profile transition happens by a tearing of its edges from the solid and the movement of the bulk fluid toward center (shown by the arrows in Fig. 8). This process represents a main point of the so-called contact angle hysteresis. 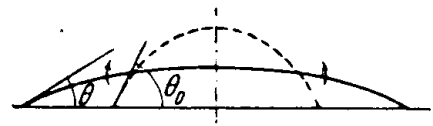

Рис. 8

The stronger the adhesion between liquid and solid, the closer $\theta$ approaches the equilibrium angle $\theta_0$, and finally the more viscous the liquid, the slower such transition takes place. For $\Delta\sigma > 2\sigma$ a thin film appears (similarly to the downward running drop) on the surface, this film is able to spread and produce small droplets surrounding the parent drop. The dynamics of such processes and a similar process of the drop rolling down an inclined surface has significant but solvable mathematical difficulties. I hope to return to questions related to this topic later.

In conclusion, we note that the motion of drops or bubbles over a solid can be determined not only by the gravitational force or by external pressure. It can involve the inertia forces (if, for example, during contact with a solid the drop has a finite velocity in tangential or normal directions, as happens for a rain drop), as well as electric forces, and so on. Based on the previous considerations, we believe that it is not, in principle, difficult to determine in which cases the drop will preserve its integrity on impact; or when the drop will separate into a set of independent droplets; or in which cases the droplets will stay on the surface and in which they will bounce up (the spray phenomena), and so forth.

Leningrad Physics Technical Institute
Academy of Science USSR